\documentclass[aps,pra,superscriptaddress,twocolumn,showpacs]{revtex4}
\usepackage{amsmath}
\usepackage{graphicx}
\usepackage{amssymb}
\usepackage{bbm, color}

\newcommand*{\cl}[1]{{\mathcal{#1}}}
\newcommand*{\bb}[1]{{\mathbb{#1}}}
\newcommand{\ket}[1]{\left|#1\right>}

\newcommand{\proj}[2]{| #1 \rangle\!\langle #2 |}
\newcommand*{\1}{{\mathbbm{1}}}
\newcommand{\T}{\mbox{$\textsf{tr}$}}

\begin{document}

\title{Gaussian private quantum channel with squeezed coherent states}

\author{Kabgyun Jeong}
\author{Jaewan Kim}
\affiliation{School of Computational Sciences, Korea Institute for Advanced Study,
  Hoegiro 85, Dongdaemun, Seoul 130-722, Korea}
\author{Su-Yong Lee}
\affiliation{Centre for Quantum Technologies,
 National University of Singapore, 3 Science Drive 2, 117543 Singapore, Singapore}

\date{\today}

\begin{abstract}
While the objective of conventional quantum key distribution (QKD) is to secretly generate and share the classical bits concealed in the form of maximally mixed quantum states, that of private quantum channel (PQC) is to secretly transmit individual quantum states concealed in the form of maximally mixed states using shared one-time pad and it is called Gaussian private quantum channel (GPQC) when the scheme is in the regime of continuous variables. We propose a GPQC enhanced with squeezed coherent states (GPQCwSC), which is a generalization of GPQC with coherent states only (GPQCo) [Phys. Rev. A 72, 042313 (2005)]. We show that GPQCwSC beats the GPQCo for the upper bound on accessible information. As a subsidiary example, it is shown that the squeezed states take an advantage over the coherent states against a beam splitting attack in a continuous variable QKD. It is also shown that a squeezing operation can be approximated as a superposition of two different displacement operations in the small squeezing regime.
\end{abstract}

\maketitle

\section{Introduction}
The notion of private quantum channel~(PQC) or quantum one-time pad~\cite{AMTW00} is very useful in quantum information processing, such as superdense coding~\cite{HHL04}, quantum data hiding~\cite{HLSW04}, quantum state sharing protocol~\cite{CJ14} (for improving their efficiency), and the proof of additivity counter-example of the classical capacity on quantum channels~\cite{HW08,H09}. The PQC is briefly introduced as follows. If the two communicating parties, Alice and Bob, share a classical secret key (e.g., via quantum key distribution procedure), then PQC can be used to transmit an arbitrary unknown quantum state from Alice to Bob securely. The intermediate state in PQC is close to the maximally mixed state, so the state exhibits almost maximum entropy. The receiver Bob always decrypts the encoded state by using the unitary inverse operations from the pre-shared secret key, whereas no third party (not having the key) can obtain the original quantum state.
Private quantum channel which belongs to a completely positive and trace preserving-map, represents the transformation of any quantum states into the maximally mixed state. It is different from the private capacity of quantum channels~\cite{PGBL09,TGW14,LWZG09} that is the maximally transmitted rate of classical secret information on quantum channels.
A discrete version of private quantum channel was first proposed by Ambainis \textit{et al}.~\cite{AMTW00} in 2000, and the optimality of PQC was proved that we need exactly $d^2$ unitary operations to encrypt a $d$-dimensional quantum state \cite{NK06,BZ07}. In the case of approximate encryption, it is sufficient to have the number of unitary operations being less than $d\log d$ \cite{HLSW04,DN06,A09}.

Then, it is natural to ask how we can realize the PQC in continuous variable~(CV) systems. Previously Br\'{a}dler proposed CV private quantum channel (PQC) using coherent states that are obtained by displacement operations on the vacuum state~\cite{B05}, where he defined a CV maximally mixed state in Gaussian regime and then constructed GPQC via the \textit{conformation} method of coherent states. Generally a single-mode Gaussian state is parametrized as a combination of displacement, squeezing operations and a thermal field~\cite{MMS02}. Specifically squeezed states, which were considered in CV quantum key distribution \cite{R00_1,R00_2,H00,GP01,CLA01}, are crucial for a security demonstration of quantum key distribution using coherent states \cite{GAWBCG03}. Moreover squeezed coherent states are useful for enhancing the security of quantum cryptography~\cite{LZO05,YHSG07}, and for improving phase sensitivities of interferometers \cite{C80}.

In this paper, we generalize the Gaussian private quantum channel~(GPQC) with a combination of displacement and \textit{squeezing} operations. Explicitly, we construct GPQC in terms of the displacement and the squeezing elements, $\exp\big[-r_p^2\{1-\tanh{r}\cdot\cos(2\theta_{pq}-\phi)\}\big]$ whereas Br\'{a}dler's GPQC is represented only by the displacement element, $\exp(-r_p^2)$. Then, we study a subsidiary example of GPQC with squeezed coherent states (GPQCwSC), especially for an eavesdropping attack. In the limit of small squeezing, furthermore, we show that the squeezed coherent states can approach a non-Gaussian regime by replacing the squeezing operation with a non-Gaussian operation, i.e., a superposition operation of two different displacements.

\section{Gaussian private quantum channel~(GPQC): coherent states} \label{bradler}
Gaussian private quantum channel (GPQC) was introduced by Br\'{a}dler in 2005, where he defined a maximally mixed state as $\1_b$ in Gaussian regime~\cite{B05}. Similarly to the discrete case (identity over the dimension: $\1/d$), the CV maximally mixed state in phase space has a broad Gaussian shape (because equiprobable mixture depends only on the radius at some boundary). Br\'{a}dler's main proposition is that the Hilbert-Schmidt distance $d_{HS}$ between the CV maximally mixed state and PQC-encryption of arbitrary coherent states is very close for sufficiently large $N$ ($N$: number of input displacement operations),
\begin{equation} \label{eq:bradler}
d_{HS}\left(\1_b,\Gamma_N\right)\approx\sqrt{N^{-2}+O(N^{-4})}\ll1,
\end{equation}
where $\Gamma_N$ denotes the mixture of all conformations of coherent states that will be defined in Eqs.~(\ref{eq:sin}) and (\ref{eq:cos}).  Also note that $d_{HS}(\rho_1,\rho_2):=\sqrt{\T(\rho_1-\rho_2)^2}$ for any matrices $\rho_{1,2}$ and it is symmetric, $d_{HS}(\rho_1,\rho_2)=d_{HS}(\rho_2,\rho_1)$. By using the unitary invariance of the distance, we can prove the statement on an arbitrary coherent state $\ket{\beta}$: for $\ket{\beta}$ and CV private quantum channel $\cl{N}_N$, $d_{HS}\big(\1_b^\beta,\cl{N}_N(\proj{\beta}{\beta})\big)=d_{HS}\left(\1_b,\Gamma_N\right)$, where $\1_b^\beta$ is a displaced CV maximally mixed state from $\1_b$ to the position of $\ket{\beta}$. The proof is a bit complex but straightforward (See details in Ref.~\cite{B05}).

Now we review the (Br\'{a}dler's) CV maximally mixed state~\cite{B05}. A CV maximally mixed state can be chosen as an integral performed over all possible single mode states within the boundary circle of radius $r\le b$ in a coherent state $\ket{\alpha}$. If $r>b$, the occurrence probability is 0. The coherent state is created by applying the displacement operator $\hat{D}(\alpha)=e^{\alpha\hat{a}^\dagger-\alpha^*\hat{a}}$ to the vacuum state $\ket{0}$ as $\ket{\alpha}=\hat{D}(\alpha)\ket{0}=e^{-|\alpha|^2/2}\sum_{n=0}^\infty\frac{\alpha^n}{\sqrt{n!}}\ket{n}$. Then, we have the CV maximally mixed state
\begin{align}
\1_b&=\frac{1}{C}\int \proj{\alpha}{\alpha}d^2\alpha \nonumber\\
&=\frac{1}{b^2}\sum_{n=0}^\infty\left(1-\sum_{k=0}^n\frac{b^{2k}}{k!}e^{-b^2}\right)\proj{n}{n},
\end{align}
where the normalization constant is $C=\pi b^2$.

The purpose of GPQC is to encrypt an input coherent state into a high entropy state. Thus the encryption should be close to the maximally mixed state in Hilbert-Schmidt distance. In order to do that, we introduce a notion of conformation through vacuum displacements. Note that $\alpha_{pq}=r_pe^{i\theta_{pq}}$ for $\theta_{pq}=\frac{\pi}{p}(2q-1)$, where $p$ and $q$ are positive integers. For some fixed $p$, an input coherent state is described by $\ket{\alpha_{pq}}=\ket{r_pe^{i\theta_{pq}}}=e^{-r_p^2/2}\sum_{m=0}^\infty\frac{(r_pe^{i\theta_{pq}})^m}{\sqrt{m!}}\ket{m}$. From the Ref.~\cite{B05}, the general and slightly modified
$p$-\textit{conformation} ($p\in\bb{Z}^+$) is given by the following equations

\begin{align}
\rho_p&=\frac{1}{p}\sum_{q=1}^p\proj{\alpha_{pq}}{\alpha_{pq}} \nonumber\\
&=\frac{e^{-r_p^2}}{p}\sum_{q=1}^p\sum_{m,n=0}^\infty\frac{r_p^{m+n}}{\sqrt{m!n!}}
e^{i\theta_{pq}(m-n)}\proj{m}{n} \nonumber \\
&=\frac{e^{-r_p^2}}{p}\sum_{m,n=0}^\infty\frac{r_p^{m+n}}{\sqrt{m!n!}}
e^{-\frac{i\pi}{p}(m-n)}\sum_{q=1}^pe^{\frac{2\pi i}{p}q(m-n)}\proj{m}{n} \nonumber \\
&=e^{-r_p^2}\sum_{m,n=0}^\infty\frac{r_p^{m+n}}{\sqrt{m!n!}}
(-1)^{\frac{m-n}{p}}\delta_{m=n(\mathrm{mod}~p)}\proj{m}{n}, \label{eq:sin} \\
&=e^{-r_p^2}\sum_{m,n=0}^\infty\frac{r_p^{m+n}}{\sqrt{m!n!}}\proj{m}{n}
\delta_{m,n(\mathrm{mod}~p)}, \label{eq:cos}
\end{align}
where $\sum_{q=1}^pe^{\frac{2\pi i}{p}q(m-n)}=p$ for $m=n~\mathrm{mod}~p$, and 0 for otherwise. The Eq.~(\ref{eq:cos}) is followed by the absorption of the phase term into $\ket{m}$'s and $\ket{n}$'s. This is equivalent to the Br\'{a}dler's original $p$-conformation. The conformation technique provides an equiprobable positioning of vacuum states at some fixed radius $r_p$, so the uniformity of the distribution of CV quantum states is strengthened.

Finally we review the mixture of all $p$-conformations ($p=1,\ldots,N$). Suppose that $N\ge1$ and define $r_p=\frac{(p-1)b}{N}\le b$, then
\begin{equation} \label{mixture}
\Gamma_N=\frac{1}{M}\sum_{p=1}^Np\rho_p
=\frac{1}{M}\sum_{p=1}^N\sum_{q=1}^p\hat{D}(\alpha_{pq})\proj{0}{0}
\hat{D}^\dagger(\alpha_{pq}),
\end{equation}
where $M=N(N+1)/2$. As an encrypted state of GPQC, the $\Gamma_N$ represents the output state of PQC over (uniformly chosen) $M$ unitary operations, where the input state is in vacuum state $\proj{0}{0}$. One of the $M$ CV states is fixed by pre-shared classical secret key between Alice and Bob as (classical) one-time pad, and then it is sent to Bob. The Br\'{a}dler's proposition states that $\Gamma_N$ is sufficiently close to the CV maximally mixed state. Encoding an arbitrary coherent state $\ket{\beta}$ is essentially equivalent to the vacuum state encryption for the unitary invariance of the distance: $d_{HS}\left(\1_b^\beta,\cl{N}_N(\proj{\beta}{\beta})\right)=d_{HS}\left(\1_b,\Gamma_N\right)$. Also note that, for any completely positive and trace-preserving (CPT) map $\cl{N}_N$, $\cl{N}_N(\proj{\beta}{\beta})=\hat{D}(\beta)\Gamma_N\hat{D}^\dagger(\beta)$ (See Eq.~(9) in Ref.~\cite{B05}). Therefore we derive Br\'{a}dler's main result as Eq.~(\ref{eq:bradler}) by combining the above properties of CV maximally mixed state, $p$-conformation, and its mixture.

\section{Gaussian private quantum channel: squeezed coherent states} \label{sgpqc}
To construct a GPQCwSC, we examine a single-mode squeezed vacuum state. A single-mode squeezing operation is defined by $\hat{S}(\xi)=\exp\left[\frac{\xi^*\hat{a}^2-\xi\hat{a}^{\dagger2}}{2}\right]$, where $\xi=re^{i\phi}$. When we apply the squeezing operator to the vacuum state, we produce a squeezed vacuum state such that
\begin{equation*}
\ket{\xi,0}:=\frac{1}{\sqrt{\cosh{r}}}\sum_{n=0}^\infty
\frac{\sqrt{(2n)!}}{2^nn!}(-e^{i\phi}\tanh{r})^n\ket{2n}.
\end{equation*}
Then, applying a displacement operation, we obtain a coherent squeezed state ($|\alpha,\xi\rangle=\hat{D}(\alpha)\hat{S}(\xi)|0\rangle$) which forms an overcomplete set, i.e., $\frac{1}{\pi}\int d^2\alpha |\alpha,\xi\rangle\langle \alpha,\xi|=\1$ \cite{BR97}. It is a main ingredient of the squeezed CV conformation.

For simplicity, we consider a squeezed coherent state, instead of the coherent squeezed state. It is reasonable that squeezed coherent states are transformed into coherent squeezed states by the relation, $\hat{S}(\xi)\hat{D}(\alpha)|0\rangle=\hat{D}(\alpha\cosh{r}-\alpha^{*}e^{i\phi}\sinh{r})\hat{S}(\xi)|0\rangle$ \cite{BR97}. Generally, a squeezed coherent state represents squeezing of a coherent state \cite{VW06},
\begin{align}
\ket{\xi,\alpha}=&\hat{S}(\xi)\hat{D}(\alpha)\ket{0} \nonumber\\
=&\frac{\left(\nu/2\cosh{r}\right)^{m/2}}{\sqrt{\cosh{r}\cdot m!}}\exp\left[-\frac{1}{2}\left(
|\alpha|^2-\frac{\nu^*\alpha^2}{\cosh{r}}\right)\right] \nonumber\\
&\times H_m\left(\frac{\alpha}{\sqrt{2\nu\cosh{r}}}\right)\ket{m}, \label{scstate}
\end{align}
where $\nu:=e^{i\phi}\sinh{r}$ and $\phi=\arg(\xi)$, the argument of the squeezing parameter $\xi$. $H_m(\cdot)$ denotes the $m$th-degree complex Hermite polynomials. By exploiting the Eq.~(\ref{scstate}), we can derive a squeezed conformations and its mixture in the following section. Then we prove that, for sufficiently large $N$ and for any squeezing of a coherent state $\ket{\beta}$, there exists a CPT map $\cl{N}$ such that $d_{HS}\left(\1_b^{(\beta,\xi)},\cl{N}_N(\xi,\proj{\beta}{\beta})\right)\ll1$. (See following second section.)

\subsection{General squeezed conformations} \label{sgpqc1}
Now, we show the explicit calculation of the squeezed $p$-conformation. Let us apply the squeezing operation to the coherent state ($\ket{\alpha_{pq}}$),
\begin{align}
\hat{S}(\xi)\ket{\alpha_{pq}}
=&\sum_{m=0}^\infty\frac{\left(\frac{\nu}{2\cosh{r}}\right)^{m/2}}{\sqrt{\cosh{r}\cdot m!}}
e^{-\frac{1}{2}\left(|\alpha_{pq}|^2-\frac{\nu^*\alpha_{pq}^2}{\cosh{r}}\right)} \nonumber\\
&\times H_m\left(\frac{\alpha_{pq}}{\sqrt{2\nu\cosh{r}}}\right)\ket{m},
\end{align}
where $\alpha_{pq}=r_pe^{i\theta_{pq}}$ and $\nu=e^{i\phi}\sinh{r}$ ($p$ and
$q$ are positive integers). Using the following relations:
$\nu^*\alpha_{pq}^2+\nu\alpha_{pq}^{*2}=r_p^2\sinh{r}\left(e^{i(2\theta_{pq}-\phi)}
+e^{-i(2\theta_{pq}-\phi)}\right)=2r_p^2\sinh{r}\cos(2\theta_{pq}-\phi)$ and
$2\sinh{r}\cdot\cosh{r}=\sinh(2r)$, then we derive the formula

\begin{align}
&\hat{S}(\xi)\proj{\alpha_{pq}}{\alpha_{pq}}\hat{S}^\dagger(\xi) \nonumber \\
=&\sum_{m,n}\frac{\left(\frac{\tanh{r}}{2}\right)^{(m+n)/2}}{\cosh{r}\sqrt{m!n!}}
e^{i{\phi}(m-n)/2}\cdot e^{-Kr_p^2} \nonumber \\
&\times H_m\left(\frac{r_pe^{i(\theta_{pq}-\frac{\phi}{2})}}{\sqrt{\sinh(2r)}}\right)
H_n\left(c.c.\right)\proj{m}{n},
\end{align}
where $K:=1-\tanh{r}\cdot\cos(2\theta_{pq}-\phi)$ and $c.c.$ denotes the complex conjugate of the argument of $H_m$. The definition of $K$ determines the position of squeezed coherent states and the squeezing angles, as shown in Fig.~1.

Then, we can find the squeezed $p$-conformation
\begin{align}
\rho_p^\xi
=&\frac{1}{p}\sum_{q=1}^p\hat{S}(\xi)\proj{\alpha_{pq}}{\alpha_{pq}}\hat{S}^\dagger(\xi) \nonumber \\
=&\sum_{m,n=0}^\infty
\kappa_{m,n}\cdot e^{-r_p^2\{1-\tanh{r}\cdot\cos(2\theta_{pq}-\phi)\}}\proj{m}{n},
\end{align}
where $\theta_{pq}=\frac{\pi}{p}(2q-1)$ and the constant $\kappa_{m,n}$ is defined by
\begin{align}
\kappa_{m,n}:=&\frac{1}{p}\sum_{q=1}^p
\frac{\left(\tanh{r}/2\right)^{(m+n)/2}}{\cosh{r}\sqrt{m!n!}}\exp\left[i\frac{\phi}{2}(m-n)\right] \nonumber\\
&\times H_m\left(\frac{r_pe^{i(\theta_{pq}-\frac{\phi}{2})}}{\sqrt{\sinh(2r)}}\right)
H_n\left(c.c.\right).
\end{align}
For some fixed squeezing $r$ and the argument $\phi$, the (complex) Hermite polynomials are orthogonal to each other for $m\neq n$ such that the value of $\kappa_{m,n}$ becomes a constant. The $\kappa_{m,n}$ converges to $r_p^{m+n}/\sqrt{m!n!}$ as $r\to0$. Therefore, the factor for some $p$, $\exp\left[-r_p^2\{1-\tanh{r}\cdot\cos(2\theta_{pq}-\phi)\}\right]$, is the main component in Eq. (9). 

\begin{figure}
\centering
\includegraphics[width=8cm]{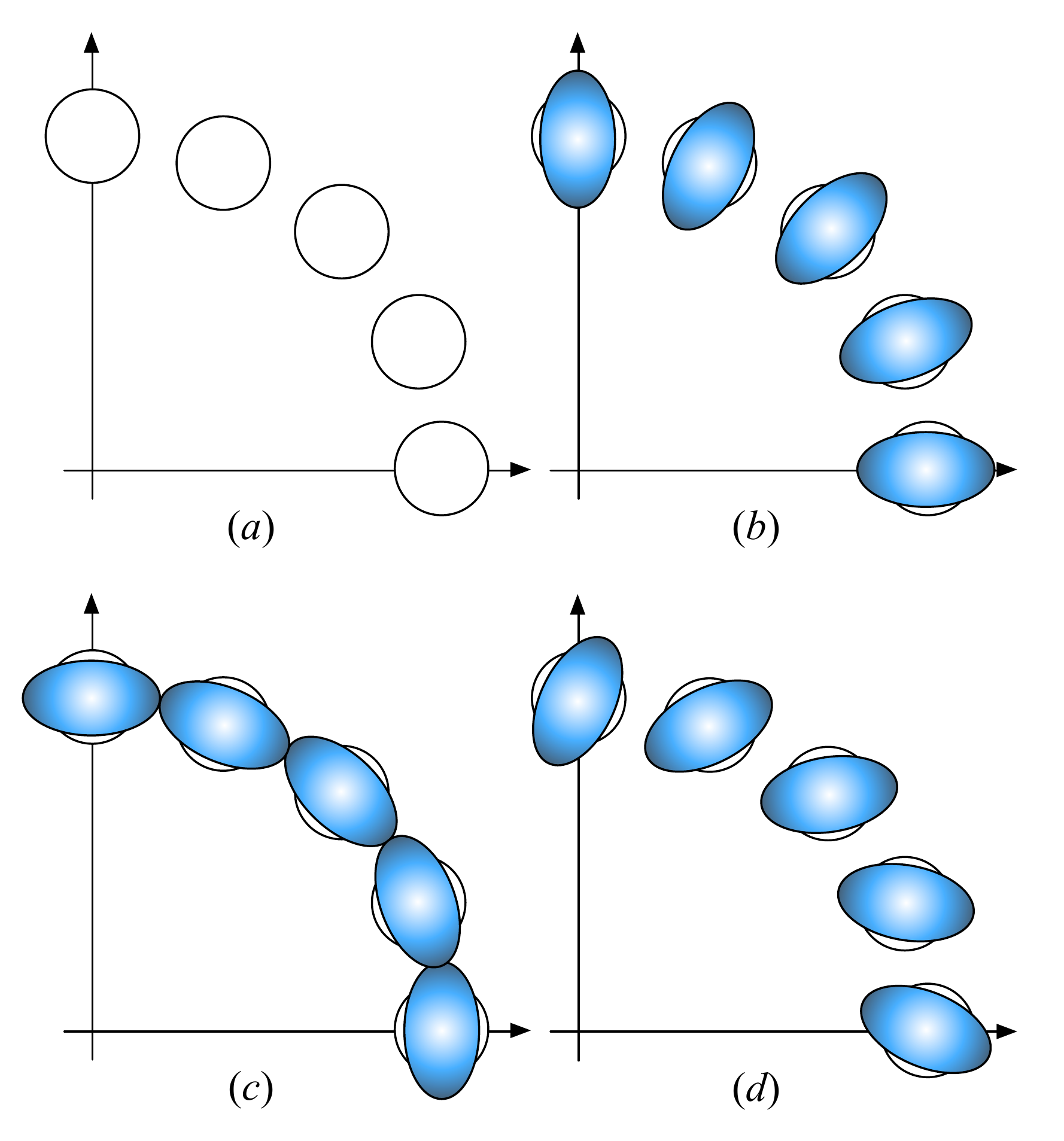}
\caption{For some fixed $r$ and $r_p$, the argument ($0\le\phi<2\pi$) of $\xi$ depends on $\theta_{pq}~\forall q\in\bb{Z}^+$. This figure represents the squeezed 16th-conformation ($r_p=r_{16}$) in the factor $K=1-\tanh{r}\cdot\cos(2\theta_{pq}-\phi)$: (a) non-squeezed ($r=0$), (b) $\phi=0$, (c) $\phi=\pm\frac{\pi}{2}$, and (d) $\phi=-\frac{\pi}{4}$ cases, respectively.}
\label{fig:Figure-1(KJ)}
\end{figure}

We finally consider the mixture of all squeezed $p$-conformations for $0\le p\le N$. Suppose $N\ge1$ and define $r_p=\frac{(p-1)b}{N}\le b$, then
\begin{equation} \label{spconf}
\Gamma_N^\xi
=\frac{1}{M}\sum_{p=1}^N\sum_{q=1}^p\hat{S}(\xi)\hat{D}(\alpha_{pq})\proj{0}{0}
\hat{D}^\dagger(\alpha_{pq})\hat{S}^\dagger(\xi),
\end{equation}
where $M=N(N+1)/2$. Alice equiprobably chooses one from the set of $M$ displacement operators $\hat{D}(\alpha_{pq})=e^{\alpha_{pq}\hat{a}^\dagger-\alpha_{pq}^*\hat{a}}$ and the squeezing parameter $r>0$. (Once again note that, for some fixed $p$ and $r$, the squeezing argument $\phi$ depends on $\theta_{pq}$ for all $q$.) Alice sends the encrypted state through a quantum channel towards Bob who performs the inverse operations to decrypt the state.

The point is that we encrypt an arbitrary input state, i.e., an arbitrary coherent state ($|\beta\rangle$). Then, we can write down a general encryption CPT map $\cl{N}$ with $M$ unitary elements as in Ref.~\cite{B05}
\begin{align*}
&\cl{N}_N(\xi,\proj{\beta}{\beta}) \nonumber\\
&=\frac{1}{M}\sum_{p=1}^N\sum_{q=1}^p\hat{S}(\xi)\hat{D}(\alpha_{pq})\hat{D}(\beta)
\proj{0}{0}\hat{D}^\dagger(\beta)\hat{D}^\dagger(\alpha_{pq})\hat{S}^\dagger(\xi) \\
&=\frac{1}{M}\sum_{p=1}^N\sum_{q=1}^p\hat{S}(\xi)\hat{D}(\beta)\hat{D}(\alpha_{pq})
\proj{0}{0}\hat{D}^\dagger(\alpha_{pq})\hat{D}^\dagger(\beta)\hat{S}^\dagger(\xi) \\
&=\hat{S}(\xi)\hat{D}(\beta)\Gamma_N\hat{D}^\dagger(\beta)\hat{S}^\dagger(\xi).
\end{align*}
From the above equation, we propose that the corresponding Hilbert-Schmidt (HS) distance is equivalent to one of Eq.~(\ref{spconf}), $d_{HS}\left(\1_b^{(\beta,\xi)},\cl{N}_N(\xi,\proj{\beta}{\beta})\right)=d_{HS}(\1_b,\Gamma_N^\xi)$ by the unitary invariance of the HS distance, whereas the states are not the same as $\cl{N}_N(\xi,\proj{\beta}{\beta})\neq\Gamma_N^\xi$.

\subsection{The proof of the main proposition and the number of secret bits} \label{sgpqc2}
Here we prove our main proposition. The proposition is as follow: For sufficiently large $N$ in any squeezing of an arbitrary coherent state $\ket{\beta}$, there exists CPT map $\cl{N}_N$ such that
\begin{align}
d_{HS}\left(\1_b^{(\beta,\xi)},\cl{N}_N(\xi,\proj{\beta}{\beta})\right)
&\le d_{HS}\left(\1_b,\Gamma_N\right) \label{lem1}\\
&\approx\sqrt{N^{-2}+O(N^{-4})}, \label{lem2}
\end{align}
where the HS distance between $\1_b$ and $\Gamma_N^\xi$ becomes quite close in sufficiently large $N$. The Eq.~(\ref{lem1}) is obtained from the unitary invariance of the HS distance. The Eq.~(\ref{lem2}) is derived via the unitary invariance of squeezing operations in the HS distance (Eq.~(\ref{sqinv}) below) and it is followed by the norm convexity (Eq.~(\ref{normconv})). Explicitly speaking, in the case of $\xi>0$, we assert that ($N\gg1$)
\begin{align}
d_{HS}\left(\Gamma_N^\xi,\Gamma_N\right)
&=d_{HS}\left(\hat{S}(\xi)\Gamma_N\hat{S}^\dagger(\xi), \Gamma_N\right) \nonumber\\
&=d_{HS}\left(\hat{S}(\xi)\proj{0}{0}\hat{S}^\dagger(\xi),\proj{0}{0}\right)\simeq0, \label{sqinv}
\end{align}
where the second equality also holds by the unitary invariance in the HS distance, i.e., for all unitary $\hat{U}:=\hat{S}\hat{D}$ and $\hat{U}':=\hat{D}\hat{S}$, $d_{HS}(\hat{U}\proj{0}{0}\hat{U}^\dagger,\hat{D}\proj{0}{0}\hat{D}^\dagger)=d_{HS}(\hat{U}'\proj{0}{0}\hat{U}'^\dagger,\hat{D}\proj{0}{0}\hat{D}^\dagger)=d_{HS}(\hat{S}\proj{0}{0}\hat{S}^\dagger,\proj{0}{0})$. In general, the last equality is not exactly equal to zero, but, asymptotically converges to 0, i.e., $d_{HS}(\hat{S}\proj{0}{0}\hat{S}^\dagger,\proj{0}{0})=\frac{2\sinh({r}/{2})}{\sqrt{\cosh r}}\simeq0$~\cite{DMMW99}.

Therefore, by using the norm convexity and the above equations (within the symmetric property of the HS distance) we derive
\begin{align}
d_{HS}(\1_b,\Gamma_N^\xi)
&\le d_{HS}(\1_b,\Gamma_N)+d_{HS}(\Gamma_N^\xi,\Gamma_N) \label{normconv} \\
&\simeq d_{HS}(\1_b,\Gamma_N)\approx(N+1)^{-1}. \nonumber
\end{align}
Thus, it implies that $\1_b$ approximately equals to the sum of the squeezed coherent states, and therefore completes the proof.

In addition, we mention the total number of unitary operations $L$ and corresponding secret bits. The number of total displacement is $M=N(N+1)/2$ and just one (pre-fixed) squeezing operation is required. From this reason, $L=M+1$. Thus, we have the number of secret bits of $\ell=\log L\sim2\log N$ for $N\gg1$. It is interesting to note that if we use the approximate random unitary channels such as in Refs.~\cite{HLSW04,DN06,A09}, then it is expected to construct PQC with only about $\frac{\ell}{2}$-bits of secret keys. There is no advantage in the key efficiency, but the accessible information can be slightly improved as follow.

\subsection{Holevo bound on the GPQCwSC} \label{accinfo}
One of important principles of the von Neumann entropy states that quantum operations never increase the quantum mutual information. By using this property, we propose that our GPQCwSC is stronger (i.e., tight upper bound) than the Br\'{a}dler's GPQC in the language of accessible information.

Formally, Br\'{a}dler's protocol~\cite{B05} with coherent states consists of a set of $\{\cl{C},p_{\cl{C}},\cl{N}_N(\proj{\beta}{\beta}),\1_b^\beta\}$, where $\cl{C}$ denotes the set of all coherent states $\ket{\beta}$, $p_{\cl{C}}$ is the probability distribution of $\cl{C}$, $\cl{N}_N(\cdot)$ is the CPT map with $N$ displacement operations, and $\1_b^\beta$ is the (displaced) CV maximally mixed state. Similarly, let us express our GPQCwSC as a set of $\{\cl{S}, p_{\cl{S}}, \cl{N}_N(\xi,\proj{\beta}{\beta}),\1_b^{(\beta,\xi)}\}$, where squeezing elements are added and the set $\cl{S}$ emphasizes the squeezing with displacement operations. Then we assert that
($\beta:=\proj{\beta}{\beta}$)
\begin{equation}
\chi\big(\{\cl{S}, p_{\cl{S}}, \cl{N}_N(\xi,\beta),\1_b^{(\beta,\xi)}\}\big)
\le\chi\big(\{\cl{C},p_{\cl{C}},\cl{N}_N(\beta),\1_b^\beta\}\big),
\end{equation}
where the Holevo information $\chi:=\max_{\cl{E}}I_{acc}(B:E)$. The $B$ and $E$ are corresponding to input and output distributions of the channel $\cl{N}_N(\cdot)$ between Bob and Eve, and the maximum of the accessible information (by Eve) is taken over all input ensemble $\cl{E}$ in the channel. Note that the quantum mutual information is defined by $I(A:B)=S(A)+S(B)-S(AB)$ for any quantum system $A$ and $B$, where $S(\sigma)=-\T\sigma\log\sigma$ is the von Neumann entropy. This fact directly comes from `the principle of quantum operation' about the entropy: For any quantum operation $Q$, $I_{acc}\big(Q(B):Q(E)\big)\le I_{acc}(B:E)$. If we substitute $Q$ to a squeezing operation $\hat{S}$, and define $B:=\hat{D}(\cl{E}_1)$ and $E:=\hat{D}(\cl{E}_2)$ for some ensembles $\cl{E}_1$ and $\cl{E}_2$, respectively, then we have
\begin{equation}
I_{acc}\big(\hat{S}(\hat{D}(\cl{E}_1):\hat{D}(\cl{E}_2))\big)
\le I_{acc}\big(\hat{D}(\cl{E}_1):\hat{D}(\cl{E}_2)\big).
\end{equation}
This provides a better upper bound on the accessible information $\chi$ than the Br\'{a}dler's analysis. In other words, the amount of eavesdropping information on the encrypted state via the GPQCwSC is less than that by the Br\'{a}dler's GPQC.

\section{Subsidiary example of GPQC} \label{applications}
We introduce a simple example that squeezed coherent states can take an advantage over coherent states in CV quantum key distribution, where the scheme is in a preliminary procedure of GPQC. To distribute quantum keys, we consider the BB84 protocol \cite{BB84}. In discrete variable systems, Alice and Bob share keys with single photon states in mutually unbiased bases. In continuous variable (CV) systems, correspondingly, Alice and Bob share keys with Gaussian states in uncertainty relation of field quadratures. Then, in the limit of small squeezing, we show that the squeezed coherent state scheme can approach even a non-Gaussian regime by replacing a squeezing operation with a superposition operation of two different displacements.

\subsection{Simple eavesdropping attack in CV quantum key distribution} \label{sQKD}
As a simple eavesdropping attack, we assume that Eve performs a beam splitting attack. As an input state, we compare a squeezed coherent state with a coherent one. For an input squeezed coherent state, Eve transforms the input state by a 50:50 beam splitter,
\begin{align}
&\hat{B}_{BE}\hat{S}_B(\xi)\hat{D}_B(\alpha)|0\rangle_B|0\rangle_E\\
&=\hat{S}_B\left(\frac{\xi}{2}\right)\hat{S}_E\left(\frac{\xi}{2}\right)\hat{S}_{BE}\left(\frac{\xi}{2}\right)\hat{D}_B\left(\frac{\alpha}{2}\right)\hat{D}_E\left(\frac{\alpha}{2}\right)|0\rangle_B|0\rangle_E,\nonumber
\end{align}
where the subscript $B$ ($E$) represents Bob (Eve), and the transformation of the squeezing operation is given by Ref.~\cite{KSBK02}. When Eve performs a measurement to get an information of the input state $(|\cdot\rangle_E)$, the state $(|\cdot\rangle_B)$ sent to Bob is disturbed by the non-local effect of the two-mode squeezing operation $\hat{S}_{BE}\big(\frac{\xi}{2}\big)$, except the uncertainty of the field quadrature. For an input coherent state, there is no non-local effect after the beam splitting attack. For the beam splitting attack, thus, Alice and Bob detect the existence of Eve much easier with the input squeezed coherent state than the input coherent one.

\subsection{Non-Gaussian regime} \label{ngaussian}
{\bf Non-Gaussian regime.} \label{ngaussian}
We show that the squeezed coherent state can approach even a non-Gaussian regime. In the limit of small squeezing, we describe a non-Gaussian regime by a truncation of the squeezing parameter, $\hat{S}(\xi)\approx 1-\frac{\xi}{2}\hat{a}^{\dag 2}+\frac{\xi^{\ast}}{2}\hat{a}^2$. However the truncation operation is not implemented by reducing the squeezing parameter in experiment. In order to apply the truncation operation to coherent states, we consider a superposition operation of two different displacements. Since an even coherent state is quite similar to a squeezed vacuum state, we derive the corresponding parameters in the limit of $r,~|\beta|^2\ll1$,
\begin{equation}
\left|\frac{\langle \beta|+\langle -\beta|}{\sqrt{2(1+e^{-2|\beta|^2})}}|\xi,0\rangle\right|^2\approx 1-|\beta|^2r\cos(2\phi-\varphi),
\end{equation}
where $\beta=|\beta|e^{i\varphi}$ and $\xi=re^{i\phi}$. When $\varphi=2\phi\pm\pi/2$, the even coherent state is approximated to the squeezed vacuum state. Note that, for $|\beta|^2\ll1$, the even coherent state is close to a Gaussian state but it is a non-Gaussian state \cite{GP10}. Therefore, the variables $r$ and $\phi$ in the squeezing parameter can be replaced by the ones $\beta$ and $\varphi$ in the even coherent state.

\begin{figure}
\centering
\includegraphics[width=8cm]{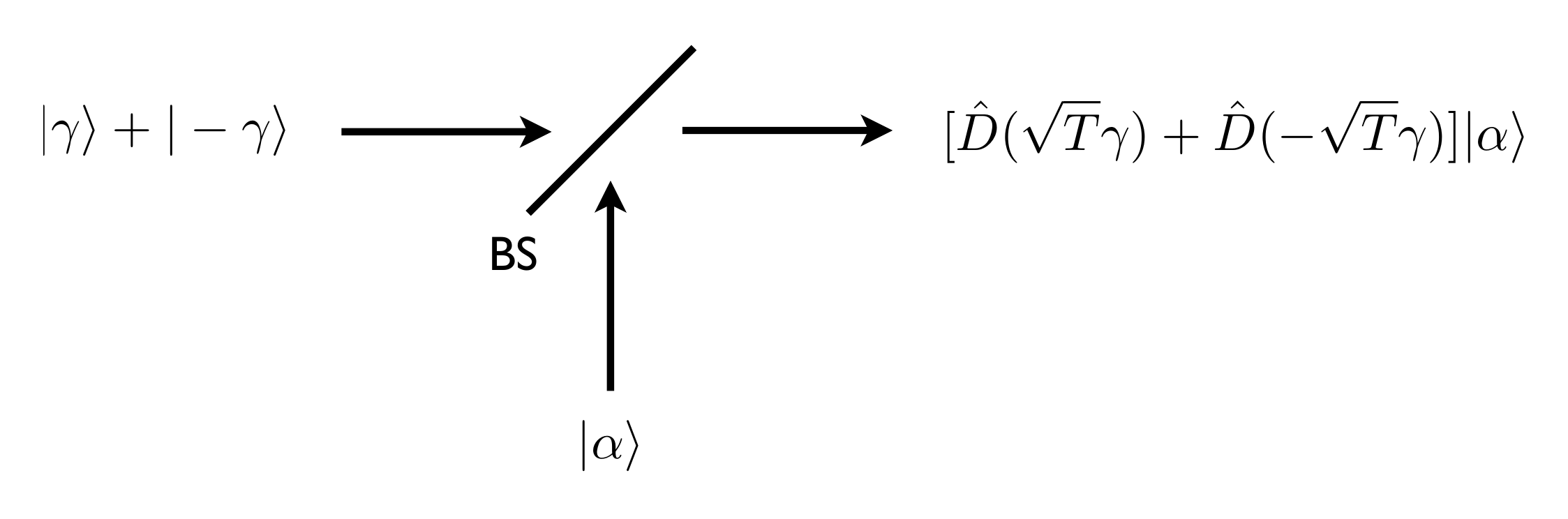}
\caption{Optical implementation for a superposition operation of two different displacements. The beam splitter is highly reflective.}
\label{fig:Figure-2(KJ)}
\end{figure}

We need to know if the uncertainties of the field quadratures are maintained by replacing the squeezed vacuum state with the even coherent state. Because CV quantum key distribution is secured via uncertainties of field quadratures \cite{R00_1,R00_2,H00}. Using the quadrature operator $\hat{X}_{\theta}=(\hat{a}e^{-i\theta}+\hat{a}^{\dag}e^{i\theta})/2$, we derive the quadrature variance of the squeezed vacuum state as
\begin{align}
\Delta X_{\theta}^2|_{sv} &=\frac{1}{4}\big[\cosh(2r)-\sinh(2r)\cos(2\theta-\phi)\big]\nonumber\\
&\approx \frac{1}{4}\big[1-2r\cos(2\theta-\phi)\big],
\end{align}
where the quadrature variance is approximated for $r\ll 1$. According to the phase parameter $\phi$, the quadrature variance oscillates between $(1-2r)/4$ and $(1+2r)/4$. The quadrature variance of the even coherent state is given by
\begin{align}
\Delta X_{\theta}^2|_{ec} &=\frac{1}{4}\left[1+\beta^2e^{-2i\theta}+\beta^{\ast 2}e^{2i\theta}+\frac{2|\beta|^2(1-e^{-2|\beta|^2})}{1+e^{-2|\beta|^2}}\right]\nonumber\\
&\approx\frac{1}{4}\big[1+2|\beta|^2\cos(2\theta-2\varphi)\big],
\end{align}
where the quadrature variance is approximated for $|\beta|^2\ll 1$. According to the phase parameter $\varphi$, the quadrature variance oscillates between $(1-2|\beta|^2)/4$ and $(1+2|\beta|^2)/4$. For the quadrature variances, thus, $|\beta|^2$ corresponds to $r$. Therefore, we find that the uncertainties of the field quadratures are maintained in the substitution of the even coherent state for the squeezed vacuum state. Note that, for the beam splitting attack, the even coherent state also plays a role of a squeezing operator by generating an entangled state with a beam splitter.

Now we see how to realize the non-Gaussian operation with an optical implementation of a superposition operation of two different displacements, as shown in Fig.~2. Previously the displacement operation was implemented by a beam splitter with high reflectivity \cite{LB02}, where the displacement amplitude is described with the multiplication ($\gamma\sqrt{T}$) of an amplitude of coherent lights ($\gamma$) and the transmission coefficient of the beam splitter ($T$). In Fig.~2, the superposition operation of two different displacements is implemented by a beam splitter with high reflectivity ($T\rightarrow 0$), where $\sqrt{T}\gamma$ represents $\beta$ in the superposition operation of two different displacements. Note that the input even coherent state can be generated by a nonlinear Kerr medium \cite{MT87,YS87,G99} in all-optical systems.

\section{Conclusion} \label{conclusion}
We have constructed GPQCwSC by an equiprobable combination of squeezed coherent states in a continuous-variable regime generalizing GPQCo and shown that GPQCwSC tightens the upper bound on accessible information. We have also presented a simple intuitive understanding of the well-known fact that the squeezed state scheme has better security than the coherent state scheme in continuous variable QKD. A class of non-Gaussian operations, superpositions of two different displacements, is shown to be an approximation of small squeezing operations. With these results, we pursue an all-optical implementation of PQC feasible with available optical technology. As some non-Gaussian states are more robust against decoherence than Gaussian states~\cite{SIS11,LKN11,NLJK12}, we look forward to investigating non-Gaussian quantum communications compared to GPQC with decoherence.

\section{acknowledgement}
We are grateful to K. Br\'{a}dler for comments. This work was partly supported by the IT R\&D program of MOTIE/KEIT [10043464]. S.Y.L. acknowledges support from FQXI and the National Research Foundation and Ministry of Education in Singapore.


\begin{thebibliography}{26}

\bibitem{AMTW00}
A. Ambainis, M. Mosca, A. Tapp, and R. de Wolf,
\textit{IEEE Symposium on Foundations of Computer Sciences}~(FOCS) p. 547 (2000).

\bibitem{HHL04}
A. Harrow, P. Hayden, and D. Leung,
\prl~\textbf{92}, 187901 (2004).

\bibitem{HLSW04}
P. Hayden, D. Leung, P.W. Shor, and A. Winter,
Commun. Math. Phys. \textbf{250}, 371 (2004).

\bibitem{CJ14}
D.P. Chi and K. Jeong,
J. Quant. Info. Sci. \textbf{4}, 64 (2014).

\bibitem{HW08}
P. Hayden and A. Winter,
Commun. Math. Phys. \textbf{284}, 263 (2008).

\bibitem{H09}
M.B. Hastings,
Nature Phys. \textbf{5}, 255 (2009).

\bibitem{PGBL09}
S. Pirandola, R. Garc\'{i}a-Patr\'{o}n, S.L. Braunstein, and S. Lloyd,
\prl~\textbf{102}, 050503 (2009).

\bibitem{TGW14}
M. Takeoka, S. Guha, and M.M. Wilde,
Nature Commun. \textbf{5}, 5235 (2014).

\bibitem{LWZG09}
K. Li, A. Winter, X. Zou, and G. Guo,
\prl~\textbf{103}, 120501 (2009).

\bibitem{NK06}
D. Nagaj and I. Kerenidis,
J. Math. Phys. \textbf{47}, 092102 (2006).

\bibitem{BZ07}
J. Bouda and M. Ziman,
J. Phys. A: Math. Theor. \textbf{40}, 5415 (2007).

\bibitem{DN06}
P.A. Dickinson and A. Nayak,
\textit{AIP Conf. Proc.} \textbf{864}, 18 (2006).
%
\bibitem{A09}
G. Aubrun,
Commun. Math. Phys. \textbf{288}, 1103 (2009).

\bibitem{B05}
K. Br\'{a}dler,
\pra~\textbf{72}, 042313 (2005).

\bibitem{MMS02}
P. Marian, T.A. Marian, and H. Scutaru, 
\prl~\textbf{88}, 153601 (2002).

\bibitem{R00_1}
T.C. Ralph,
\pra~\textbf{61}, 010303(R) (2000).

\bibitem{R00_2}
T.C.Ralph,
\pra~\textbf{62}, 062306 (2000).

\bibitem{H00}
M. Hillery, 
\pra~\textbf{61}, 022309 (2000).

\bibitem{GP01}
D. Gottesman and J. Preskill, 
\pra~\textbf{63}, 022309 (2001).

\bibitem{CLA01}
N.J. Cerf, M. L\'evy, and G. Van Assche, 
\pra~\textbf{63}, 052311 (2001).

\bibitem{GAWBCG03}
F. Grosshans, G. Van Assche, J. Wenger, R. Brouri, N.J. Cerf, and Ph. Grangier, 
Nature~\textbf{421}, 238 (2003).

\bibitem{LZO05}
Y.J. Lu, L. Zhu, and Z.Y. Ou,
\pra~\textbf{71}, 032315 (2005).

\bibitem{YHSG07}
Z.-Q. Yin, Z.-F. Han, F.-W. Sun, and G.-C. Guo,
\pra~\textbf{76}, 014304 (2007).

\bibitem{C80}
C.M. Caves, 
\prd~\textbf{23}, 1693 (1980).


\bibitem{BR97}
S.M. Barnett and P.M. Radmore,
Methods in Theoretical Quantum Optics,
Oxford University Press (1997).

\bibitem{VW06}
W. Vogel and D.-G. Welsch,
{Quantum Optics},
WILEY-VCH Verlag GmbH \& Co. KGaA (2006).

\bibitem{DMMW99}
V.V. Dodonov, O.V. Man'ko, V.I. Man'ko, and A. W\"{u}nsche,
Phys. Scr. \textbf{59}, 81 (1999).

\bibitem{BB84}
C.H. Bennett and G. Brassard,
Proceedings of IEEE International Conference on Computers, Systems and Signal Processing, \textbf{175}, 8 (1984) New York.

\bibitem{KSBK02}
M.S. Kim, W. Son, V. Bu\u zek, and P.L. Knight,
\pra~\textbf{65}, 032323 (2002).

\bibitem{GP10}
M.G. Genoni and M.G.A. Paris,
\pra~\textbf{82}, 052341 (2010).

\bibitem{LB02}
A.I. Lvovsky and S.A. Babichev,
\pra~\textbf{66}, 011801(R) (2002).

\bibitem{MT87} 
A. Mecozzi and P. Tombesi, 
\prl~\textbf{58}, 1055 (1987).

\bibitem{YS87}
B. Yurke and D. Stoler,
\pra~\textbf{35}, 4846 (1987).

\bibitem{G99}
C.C. Gerry,
\pra~\textbf{59}, 4095 (1999).

\bibitem{SIS11}
K.K. Sabapathy, J.S. Ivan, and R. Simon, 
\prl~\textbf{107}, 130501 (2011).

\bibitem{LKN11}
J. Lee, M. S. Kim, and H. Nha, 
\prl~\textbf{107}, 238901 (2011).

\bibitem{NLJK12}
H. Nha, S.-Y. Lee, S.-W. Ji, and M.S. Kim,
\prl~\textbf{108}, 030503 (2012).



\end{thebibliography}
\end{document}